\documentclass[conference]{IEEEtran}
\usepackage[pdftex]{graphicx}
\usepackage{comment}
\usepackage{booktabs}
\usepackage{threeparttable}
\usepackage{multirow}
\usepackage{multicol}
\usepackage{pifont}
\usepackage{caption}
\usepackage{subcaption}
\usepackage{color}
\usepackage{framed}
\usepackage{booktabs}
\usepackage{graphicx}
\usepackage{amsmath,amssymb,amsfonts}
\usepackage{dirtytalk}
\usepackage{hyperref}

\usepackage{cite}
\usepackage{amsmath,amssymb,amsfonts}
\usepackage{graphicx}
\usepackage{booktabs}
\usepackage{url}
\usepackage{algorithmic}
\usepackage{subcaption}
\usepackage{color}
\usepackage{framed}
\usepackage{comment}
\usepackage{hyperref}

\newcounter{t0d0_counter}
\newcommand{\nofixme}[1]{
}
\newcommand{\fixme}[1]{
 \stepcounter{t0d0_counter}
 \definecolor{shadecolor}{rgb}{1,1,0} 
 \begin{shaded}
 T0D0 \arabic{t0d0_counter}: #1
 \end{shaded}
}

\usepackage[switch]{lineno}

\begin{document}


%
\title{Towards a Unified Cybersecurity Testing Lab for Satellite, Aerospace, Avionics, Maritime, Drone (SAAMD) technologies and communications}


\author{
\IEEEauthorblockN{Andrei Costin, Hannu Turtiainen, Syed Khandker, Timo H\"{a}m\"{a}l\"{a}inen}
\IEEEauthorblockA{Faculty of Information Technology \\
University of Jyv\"{a}skyl\"{a} \\
Finland \\ 
\{ancostin,turthzu,syibkhan,timoh\}@jyu.fi}
}


%


\IEEEoverridecommandlockouts
\makeatletter\def\@IEEEpubidpullup{6.5\baselineskip}\makeatother
\IEEEpubid{\parbox{\columnwidth}{
    \textit{AUTHORS' PREPRINT, DRAFT ACCEPTED BY REVIEWERS} of Workshop on Security of Space and Satellite Systems (SpaceSec)\\
    Network and Distributed System Security (NDSS) Symposium 2023\\
    28 February - 4 March 2023, San Diego, CA, USA\\
    ISBN 1-891562-83-5\\
    https://dx.doi.org/10.14722/ndss.2023.23xxx\\
    www.ndss-symposium.org
}
\hspace{\columnsep}\makebox[\columnwidth]{}}

\maketitle

\begin{abstract}

Aviation, maritime, and aerospace traffic control, radar, communication, and software technologies received increasing attention in the research literature over the past decade, as software-defined radios have enabled practical wireless attacks on communication links previously thought to be unreachable by unskilled or low-budget attackers. 
Moreover, recently it became apparent that both offensive and defensive cybersecurity has become a strategically differentiating factor for such technologies on the war fields (e.g., Ukraine), affecting both civilian and military missions regardless of their involvement. 
However, attacks and countermeasures are usually studied in simulated settings, thus introducing the lack of realism or non-systematic and highly customized practical setups, thus introducing high costs, overheads, and less reproducibility. 
Our \emph{``Unified Cybersecurity Testing Lab''} seeks to close this gap by building a laboratory that can provide a systematic, affordable, highly-flexible, and extensible setup. 

In this paper, we introduce and motivate our \emph{``Unified Cybersecurity Testing Lab for Satellite, Aerospace, Avionics, Maritime, Drone (SAAMD)''} technologies and communications, as well as some peer-reviewed results and evaluation of the targeted threat vectors. 
We show via referenced peer-reviewed works that the current modules of the lab were successfully used to realistically attack and analyze air-traffic control, radar, communication, and software technologies such as ADS-B, AIS, ACARS, EFB, EPIRB and COSPAS-SARSAT. 
We are currently developing and integrating support for additional technologies (e.g., CCSDS, FLARM), and we plan future extensions on our own as well as in collaboration with research and industry. 
Our \emph{``Unified Cybersecurity Testing Lab''} is open for use, experimentation, and collaboration with other researchers, contributors and interested parties. 

\end{abstract}

\section{Introduction}
\label{sec:intro}

Aviation, maritime, and aerospace traffic control, radar, communication, and software technologies received increasing attention in the research literature over the past decade, as software-defined radios have enabled practical wireless attacks on communication links previously thought to be unreachable by unskilled or low-budget attackers. 
Critical protocols and implementations in these domains have been demonstrated to be either insecure or exploitable under various attacks -- EPIRB and CCSDS (Section~\ref{sec:results-sat} and \textit{our other SpaceSec23 submission on ``COSPAS-SARSAT/EPIRB''}), ADS-B~\cite{costin2012ghost,strohmeier2014realities}, AIS~\cite{balduzzi2014security,9733358}, ACARS~\cite{smith2016security}, GDL90~\cite{turtiainen2022gdl90fuzz}. 
Moreover, recently it became apparent that both offensive and defensive cybersecurity has become a strategically differentiating factor for such technologies on the war fields (e.g., Ukraine), affecting both civilian and military missions regardless of their involvement. 
However, attacks and countermeasures are usually studied in simulated settings, thus introducing the lack of realism or non-systematic and highly customized practical setups, thus introducing high costs, overheads, and less reproducibility. 

At the same time, satellite, space and aerospace is strongly interconnected with aviation, maritime and Search-and-Rescue (SAR) domains, e.g., satellites processing aviation (ACARS, ADS-B) and maritime (AIS) data arriving over various communication links. 
Given this tight interconnect of technologies, the ``additive complexity'' may give rise to additional attacks such as Cross-Channel (XC) as both theorized and demonstrated by~\cite{turtiainen2022gdl90fuzz,juvonen2022apache}, and somewhat equivalent of Cross-Channel Scripting (XCS) for IoT and web domains~\cite{bojinov2009xcs}. 

Our \emph{``Unified Cybersecurity Testing Lab''} seeks to close this gap by building a laboratory (with its associated extensible programmatic platform and testbed devices) that can provide a systematic, affordable, highly-flexible, and extensible setup. 
Consequently, our unified lab approach allows to experiment with and test the scenarios that would be otherwise hard or impossible to test in labs dedicated solely to specific domains, e.g., avionics-only, maritime-only, space-only.

\subsection{Contributions}
\label{sec:motiv}

In this paper, we introduce and detail a \emph{``Unified Cybersecurity Testing Lab for Satellite, Aerospace, Avionics, Maritime, Drone (SAAMD)''} technologies and communications, as well as some peer-reviewed results and evaluation of the targeted threat vectors. 
We show via referenced peer-reviewed works that the first modules of the lab were successfully used to realistically attack and analyze traffic control, radar, communication, and software technologies related to satellites, space, aerospace, avionics,  and maritime systems (e.g., EPIRB, CCSDS, ADS-B, AIS, ACARS). 
We are currently developing and integrating support for additional technology (e.g., drones -- FLARM, RemoteID), and we plan future extensions and improvements to our lab (e.g., GPS attacks/controls, more sophisticated environment control). 
With this, we aim to convince that a unified lab (with strong focus on space and satellite technologies) is not only beneficial but many times necessary in order to test complex scenarios as well as to be prepared for the leading role of space/satellites in years to come.

\subsection{Organization}
\label{sec:org}

The rest of this paper is organized as follows. 
We discuss related studies in Section~\ref{sec:related}.  
We describe our lab and pentesting platform in Section~\ref{sec:lab}. 
Then, in Section~\ref{sec:results}, we describe different attacking scenarios, their impact on ADS-B, ACARS, and AIS receivers, and analysis of the results. 
Finally, we conclude this paper with Section~\ref{sec:concl}.


\section{Related Work}
\label{sec:related}

Although avionics and maritime communication has been the subject of profound research~\cite{costin2012ghost,Matthias2013,strohmeier2014realities,9133434,222,6940209,manesh2018preliminary,eskilsson2020demonstrating,lundberg2014security,lundberg2014security-phd,MCCALLIE201178,Naima,Donald,crc,balduzzi2014security,ray2015deais,ray2016methodology,khandker2021cybersecurity,turtiainen2022gdl90fuzz,9473225}, the focus has been mainly on a specific attack or in theory. Recently Strohmeier et al.~\cite{strohmeier2022building} researched building an avionics laboratory for cybersecurity testing. Their approach was testbed with ``certifiable realism,'' meaning that the equipment must be capable of in-plane use. They also maintain that the testbed should be device manufacturer agnostic and in-laboratory contained. They utilized Garmin GTN 759 flight management system, Garmin GTX 3000 aircraft transponder, and Garmin GTS 8000 TCAS collision avoidance systems in their laboratory with some auxiliary equipment such as software-defined radios and Faraday cages. Currently, they support ARINC 429 avionics communication bus, secondary surveillance radar (SSR), Automatic Dependent Surveillance-Broadcast (ADS-B), Global Navigation Satellite Systems (GNSS), Airborne Collision Avoidance System (ACAS), and Traffic Alert and Collision Avoidance System (TCAS) technologies. The authors' initial tests of their testbed were successful, and the results were promising. For further research, the authors provided some guidance in their work. They concluded that constructing environment ``realism'' in a laboratory setting has trade-offs in complexity and cost, affecting the laboratory's expandability and future-proofing. They also received pushback from avionics manufacturers for collaboration and acquiring the equipment. In conclusion, Strohmeier et al.~\cite{strohmeier2022building} built a highly-capable and effective laboratory setup for trustworthy avionics cybersecurity testing.

Avionics laboratories can also be particular for thoroughly testing specific equipment. For example, they required a sophisticated testing suite when South Korea's defense department rolled out their new utility helicopter with a new kind of Mission Equipment Package (MEP) integrated mission control system. Kim et al.~\cite{kim_integrationlab} conducted a requirement assessment and designed a system integration laboratory to verify the MEP's capabilities and functionality before accepting the technology for active duty. Viana Sanchez and Taylor~\cite{viana2010reference} introduced a Reference Architecture System Testbed for Avionics (RASTA). Their goal was to define an architecture for a laboratory that could combine, at the time, the latest agreements of avionics communication as well as the requirements for end-to-end spacecraft communication.
Dey et al.~\cite{dey2018drone} investigated drone security vulnerabilities. Their testbed contained two drones (DJI Phantom 4 Pro and Parrot Bebop 2), a LabSat GPS simulator, and two mobile phones for hosting the drone-controlling application. By performing several attacks, such as deauthentication, GPS spoofing, unauthorized file access, and others, they concluded that the vulnerabilities in drones could lead to invasions of privacy, concerns with aircraft safety, and even personal injury.

Our work relates closest to the recent work of Strohmeier et al.~\cite{strohmeier2022building}. Even though our labs have inherent fundamental similarities in their designs, goals, and protocols, there are also several unique differentiating features.
First, our lab and platform already cover aerospace (EPIRB, CCSDS) and maritime (AIS), in addition to the focused aviation/avionics (ADS-S, ACARS) field. 
Second, our lab, platform, and tests crucially focus on the attacker's perspective (e.g., attack vectors, successful exploitation, new attack techniques) with subsequent defensive improvements to the affected systems, even though testing the adherence to functional specifications and cybersecurity standards is also within the scope and capabilities of our lab. 
%
Last but not least, our lab aims is fitted to research offensive and defensive cybersecurity in highly complex end-to-end scenarios. One example is researching the effect of ADS-B/AIS attacks when ADS-B/AIS is attacked directly via an interface on aircraft/ships or via interfaces on satellites supporting these links. Another example is researching the effect of attacks when the attacker pivots across protocols (e.g., ADS-B to CCSDS and vice-versa) or across devices (e.g., ADS-B transponder of aircraft to satellite RF boards vice-versa). However, another example is researching the effect of common IT vulnerabilities (e.g., log4j) in cases when vulnerable components are used in aviation, maritime, aerospace infrastructure, and devices~\cite{juvonen2022apache}.

\section{Our Testing Lab}
\label{sec:lab}

\begin{figure}[!t]
\centering
\includegraphics[width = 0.90\columnwidth, trim = {0.5in 4.5in 4.5in 3.6in}]{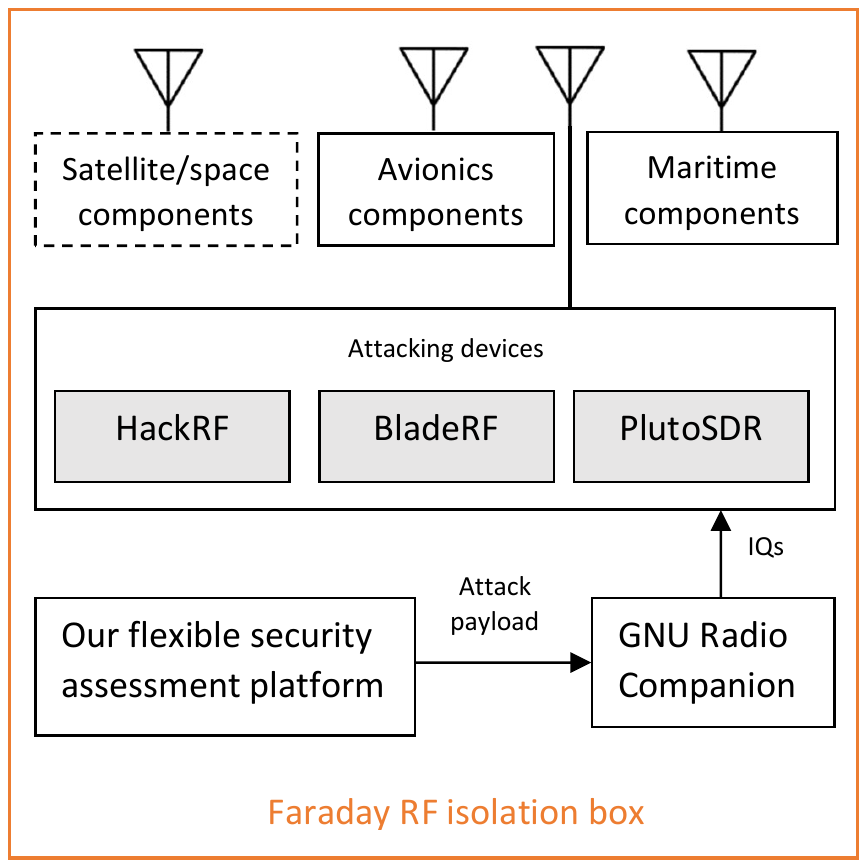}
\caption{Diagram of our lab testing approach.} 
\label{fig:setup}
\end{figure}
The main goal of our investigation was to keep the testing scenario close to the realistic one. Therefore, we used transmission-capable SDRs to generate real-like but fake or non-standard signals, e.g., for ADS-B, AIS, and ACARS. 
We then tested different avionics and maritime devices over the wireless interface. There was an air gap between the transmitters and receivers. Fig~\ref{fig:setup} shows the design of our lab. 
We developed a flexible and extensible security assessment platform that, at present, can create ADS-B 1090ES, UAT978, and AIS payload according to the protocol specifications and attack/test specifications. 
We used a signal processing software called GNU radio companion (GRC) to generate IQs of the RF signal of the payload. Then the IQs were sunk into the transmission-capable SDRs to create ADS-B and AIS RF signals. Even though one type of SDR is enough for the test, we tested three to check the attacking devices' availability. The list of different hardware and software in our laboratory is as follows.

\subsection{Software}

An RF testing laboratory requires much software to be functional. Our laboratory already employs an extensive software suite, and we are constantly adding more. In Table~\ref{tab:software}, we disclose the software we currently use at the time of writing.

\begin{table}[!h]
\caption{List of different software}
\resizebox{\columnwidth}{!}{
\begin{tabular}{|l|l|l|}
\hline
\multicolumn{1}{|c|}{\textbf{Platform}} & \multicolumn{1}{c|}{\textbf{Software name}} & \multicolumn{1}{c|}{\textbf{Functionality}} \\ \hline
\multirow{18}{*}{Aviation}     & Dump1090            & Decoding and displaying 1090ES data                \\ \cline{2-3} 
                               & Dump978             & Decoding and displaying UAT978 data                \\ \cline{2-3} 
                               & RTL1090             & Decoding and displaying 1090ES data                \\ \cline{2-3} 
                               & PlanePlotter        & Displaying 1090ES data                             \\ \cline{2-3} 
                               & Micro ADS-B         & Displaying 1090ES data                             \\ \cline{2-3} 
                               & QGround Control     & Displaying 1090ES data                             \\ \cline{2-3} 
                               & Mission Planner     & Displaying 1090ES data                             \\ \cline{2-3} 
                               & Garmin Pilot        & Displaying 1090ES and UAT978 data                  \\ \cline{2-3} 
                               & ForeFlight          & Displaying 1090ES and UAT978 data                  \\ \cline{2-3} 
                               & Airmate             & Displaying 1090ES and UAT978 data                  \\ \cline{2-3} 
                               & AvPlan              & Displaying 1090ES and UAT978 data                  \\ \cline{2-3} 
                               & Easy VFR4           & Displaying 1090ES and UAT978 data                  \\ \cline{2-3} 
                               & FlyQ                & Displaying 1090ES and UAT978 data                  \\ \cline{2-3} 
                               & Stratus Insight     & Displaying 1090ES and UAT978 data                  \\ \cline{2-3} 
                               & OZRunways           & Displaying 1090ES and UAT978 data                  \\ \cline{2-3} 
                               & Horizon             & Displaying 1090ES data                             \\ \cline{2-3} 
                               & SkyDemon            & Displaying 1090ES and UAT978 data                  \\ \cline{2-3} 
                               & ADL Connect         & Displaying 1090ES data                             \\ \hline
\multirow{11}{*}{Maritime}     & OpenCPN             & Displaying AIS data                                      \\ \cline{2-3} 
                               & iRegatta            & Displaying AIS data                                      \\ \cline{2-3} 
                               & Ships               & Displaying AIS data                                      \\ \cline{2-3} 
                               & Boating             & Displaying AIS data                                      \\ \cline{2-3} 
                               & iBoating            & Displaying AIS data                                      \\ \cline{2-3} 
                               & Boat Beacon         & Displaying AIS data                                      \\ \cline{2-3} 
                               & AF track            & Displaying AIS data                                      \\ \cline{2-3} 
                               & RTL AIS driver      & Decoding AIS data                                        \\ \cline{2-3} 
                               & AIS Share           & Sharing AIS data                                         \\ \cline{2-3} 
                               & ShipPlotter         & Decoding and displaying AIS data                         \\ \cline{2-3} 
                               & AISmon              & Decoding and sharing AIS signal                          \\ \hline
\multirow{3}{*}{Others} & SDR Sharp           & Receiving RF signal                                      \\ \cline{2-3} 
                               & GNU Radio Companion & Generating IQs  \\ \cline{2-3} 
                               & Our Pentesting Platform  & Generating offensive/non-standard payloads                         \\ \hline
\end{tabular}
}
\label{tab:software}
\end{table}

\subsection{Avionics components}
We tested 11 ADS-B receivers, and some had transmitting capability too. They all support ADS-B 1090ES, four support dual ADS-B mode, and four support UAT978. Table~\ref{tab:aviation} shows our laboratory's avionics components.

\begin{table}[!h]
\caption{List of Avionics components in our laboratory}
\resizebox{\columnwidth}{!}{
\begin{tabular}{|l|l|}
\hline
\multicolumn{1}{|c|}{\textbf{Device name}} & \multicolumn{1}{c|}{\textbf{Functionality}} \\ \hline
uAvionix Skyecho2         & 1090ES and UAT978 receiver. 1090ES transmitter \\ \hline
uAvionix echoUAT          & 1090ES and UAT978 receiver. UAT978 transmitter \\ \hline
ForeFlight Sentry         & 1090ES and UAT978 receiver                     \\ \hline
Garmin GDL 52             & 1090ES and UAT978 receiver                     \\ \hline
Aerobits TR-1W            & 1090ES receiver and transmitter                \\ \hline
ADL 180                   & 1090ES receiver                                \\ \hline
Helios Avionics SensorBox & 1090ES receiver                                \\ \hline
Plane Gadget Radar (PGR)  & 1090ES receiver                                \\ \hline
Aerobits EVAL-TT-SF1      & 1090ES receiver                                \\ \hline
PX4                       & 1090ES receiver                                \\ \hline
Cube Orange               & 1090ES receiver                                \\ \hline
\end{tabular}
}
\label{tab:aviation}
\end{table}

\begin{figure}[!h]
\centering
\includegraphics[width = 1.00\columnwidth, trim = {0.0in 0.0in 0.0in 0.0in}]{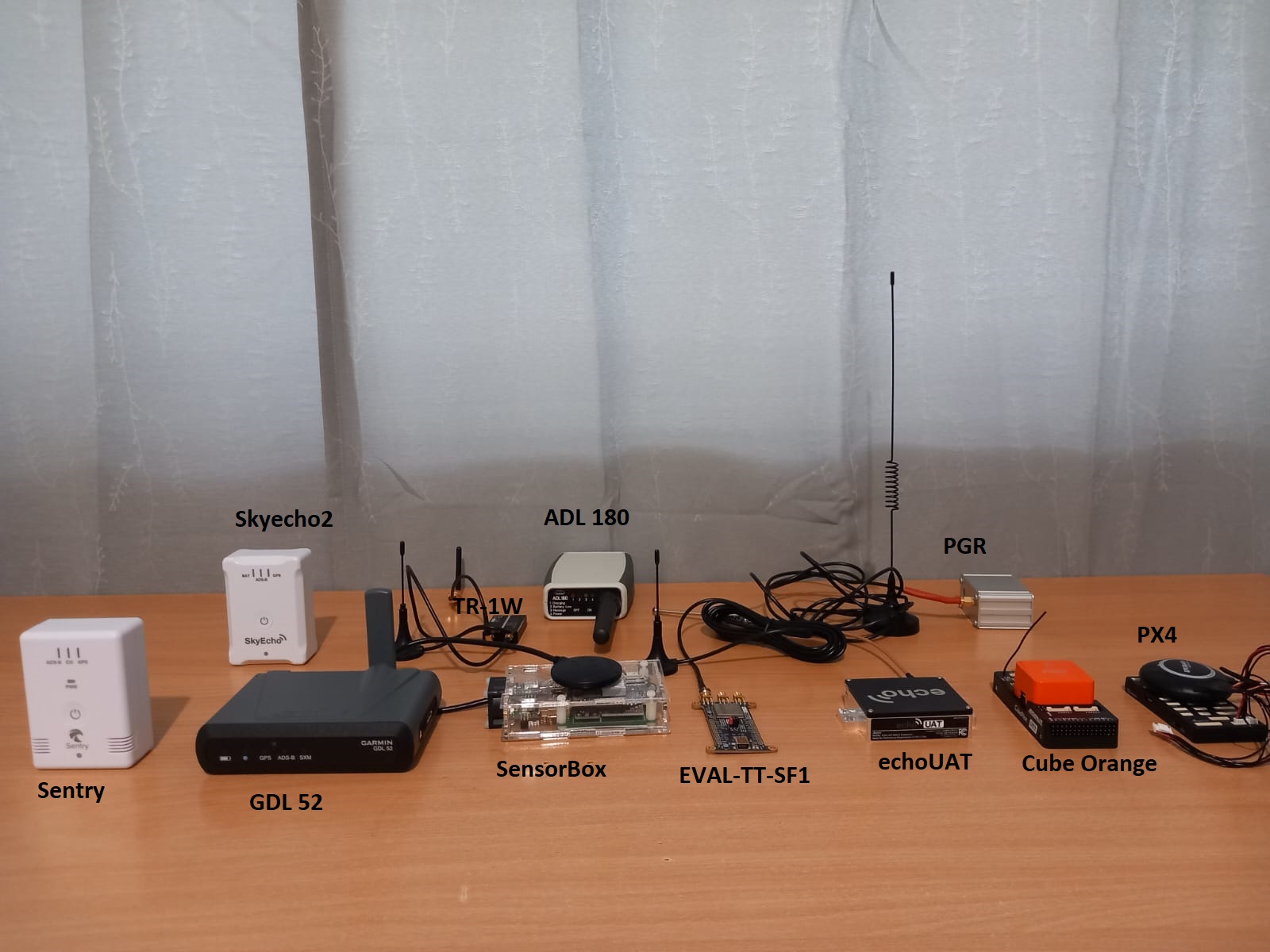}
\caption{Aviation/avionics components} 
\label{fig:avionics}
\end{figure}

Figure~\ref{fig:avionics} shows the avionics component of our laboratory.

\subsection{Maritime components}
We tested a commercial transponder, a professional AIS receiver, and many RTL SDR-based mobile AIS setups in our laboratory. Table~\ref{tab:maritime} shows the list.

\begin{table}[!h]
\caption{List of maritime components in our laboratory}
\resizebox{\columnwidth}{!}{
\begin{tabular}{|l|l|}
\hline
\multicolumn{1}{|c|}{\textbf{Device name}} & \multicolumn{1}{c|}{\textbf{Functionality}} \\ \hline
Matsutec HP-33A    & Stand alone AIS transponder              \\ \hline
Quark-elec QK-A027 & AIS receiver                             \\ \hline
McMurdo G8 & COSPAS-SARSAT AIS/EPIRB transmitter              \\ \hline
RTL-SDR            & RF front-end for AIS mobile applications \\ \hline
\end{tabular}
}
\label{tab:maritime}
\end{table}

\begin{figure}[!h]
\centering
\includegraphics[width = 1.00\columnwidth, trim = {0.0in 0.0in 0.0in 0.0in}]{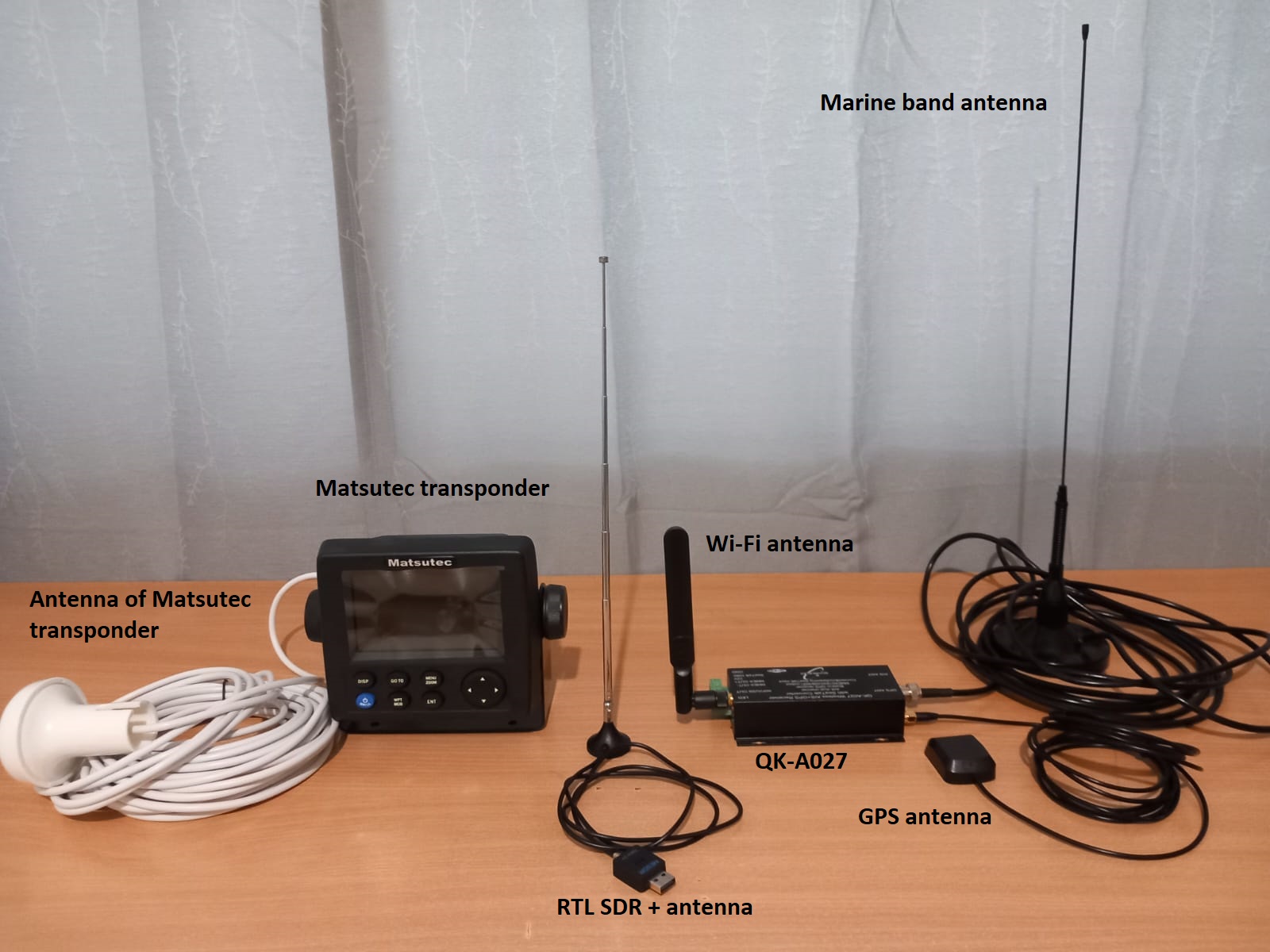}
\caption{Maritime components} 
\label{fig:maritime}
\end{figure}

Figure~\ref{fig:maritime} shows the maritime component of our laboratory.

\subsection{Satellite and aerospace components}

\begin{figure}[!h]
\centering
\includegraphics[width = 1.00\columnwidth]{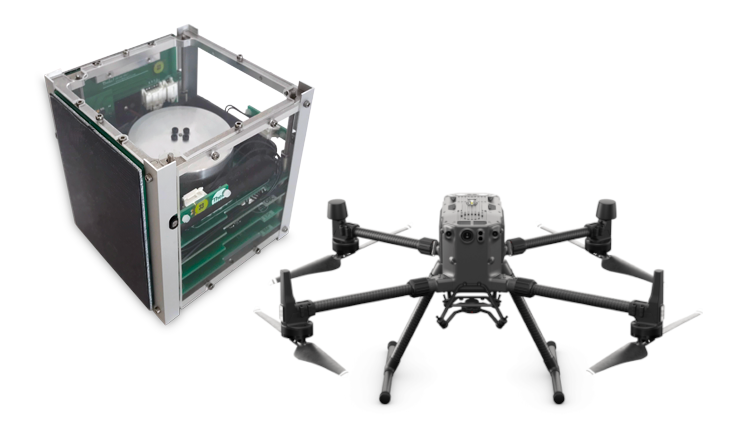}
\caption{Satellite/aerospace and drone components -- Theia Space ESAT and DJI MATRICE 300 RTK} 
\label{fig:lab_gear1}
\end{figure}

\begin{figure}[!htb]
\centering
\includegraphics[width=0.60\columnwidth]{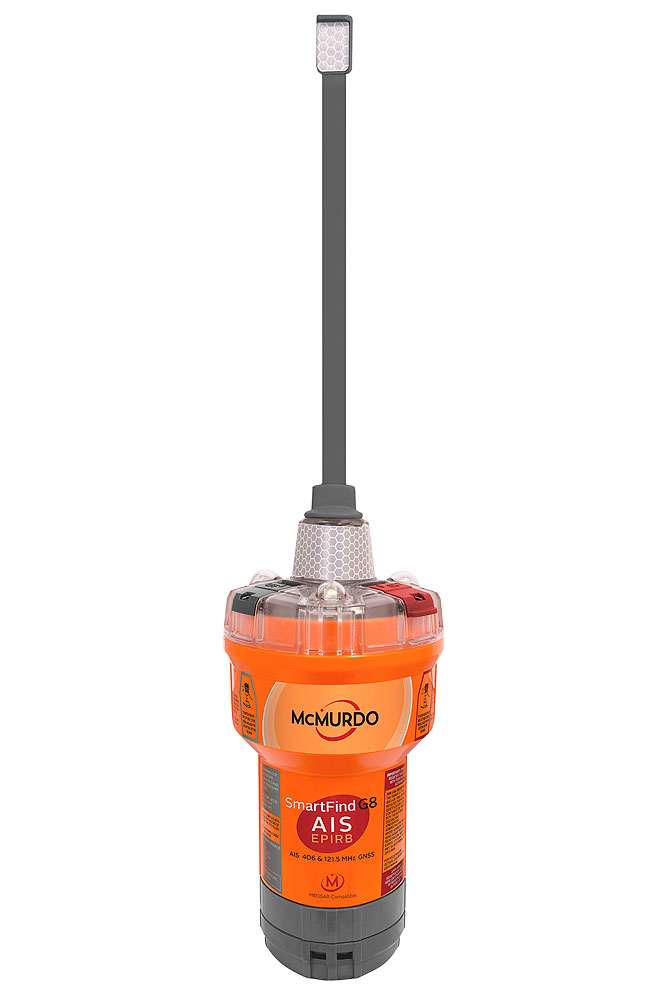}
\caption{Satellite/aerospace technology -- a COSPAS-SARSAT EPIRB McMurdo G8~\cite{mcmurdo-g8}}
\label{fig:mcmurdo}
\end{figure}

At the time of this writing, we already have in our lab a space device (Theia Space ESAT), COSPAS-SARSAT devices, and an aerospace drone device (DJI MATRICE 300 RTK), as depicted in Figures~\ref{fig:lab_gear1}~\ref{fig:mcmurdo}. 
A fast preliminary implementation already allowed us to discover some Denial-of-Service vulnerabilities on the satellite device that effectively disables RF/COMM communication board and requires a hard reboot. 
Thanks to our lab and platform, we discovered this is a problematic scenario for satellites, as availability is a top priority in the field. 
Immediate future work is to research, develop, and integrate into our pentesting platform additional and complete support for software and protocols for these devices and subsequently thoroughly evaluate their cybersecurity posture when facing both existing attacks~\cite{costin2012ghost,khandker2021cybersecurity,turtiainen2022gdl90fuzz,9749067,9733358} and perhaps novel ones. 

\begin{table}[!h]
\caption{List of satellite, space, aerospace devices}
\resizebox{\columnwidth}{!}{
\begin{tabular}{|l|l|}
\hline
\multicolumn{1}{|c|}{\textbf{Device name}} & \multicolumn{1}{c|}{\textbf{Functionality}} \\ \hline

Theia Space ESAT    & a) CCSDS receiver; b) ``System security'' payloads/boards \\ \hline

DJI Matrice 300 RTK   & a) ADS-B, FLARM, RemoteID; b) Remote-carrying of ``attacking devices''~\ref{sec:attack-dev}  \\ \hline

McMurdo G8      &   COSPAS-SARSAT EPIRB 406 transmitter \\ \hline

\end{tabular}
}
\label{tab:attack}
\end{table}

\subsection{Attacking devices}
\label{sec:attack-dev}
We used three types of SDRs to transmit the attack/test signals. All of them supported sending of ADS-B signals. HackRF and BladeRF support AIS transmission, but the Pluto SDR's operating frequency is out of the AIS frequency range. Table~\ref{tab:attack} shows the list of the attacking devices.

\begin{table}[!h]
\caption{List of attacking devices}
\resizebox{\columnwidth}{!}{
\begin{tabular}{|l|l|}
\hline
\multicolumn{1}{|c|}{\textbf{Device name}} & \multicolumn{1}{c|}{\textbf{Functionality}} \\ \hline
HackRF    & Generating ADS-B, AIS, EPIRB signals using Python/GRC \\ \hline
BladeRF   & Generating ADS-B, AIS, EPIRB signals using Python/GRC \\ \hline
Pluto SDR & Generating ADS-B, EPIRB signals using Python/GRC      \\ \hline
\end{tabular}
}
\label{tab:attack}
\end{table}

\begin{figure}[!h]
\centering
\includegraphics[width = 1.00\columnwidth, trim = {0.0in 0.0in 0.0in 0.0in}]{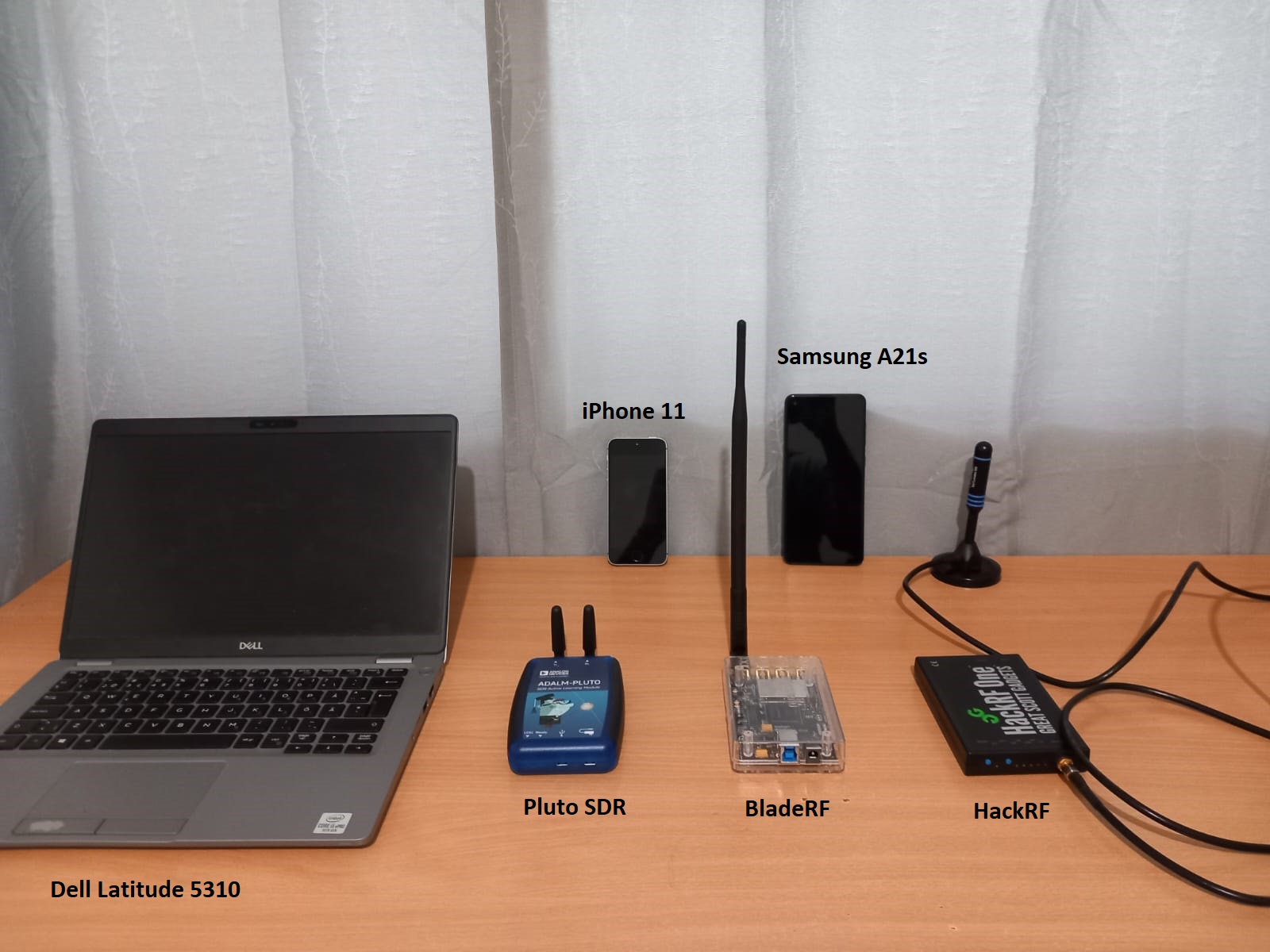}
\caption{Attacking SDRs of our laboratory with auxiliary devices} 
\label{fig:attacking}
\end{figure}

Figure~\ref{fig:attacking} shows the attacking and auxiliary devices of our laboratory.

\subsection{Environment control components}

\begin{figure}[!h]
\centering
\includegraphics[width = 1.00\columnwidth]{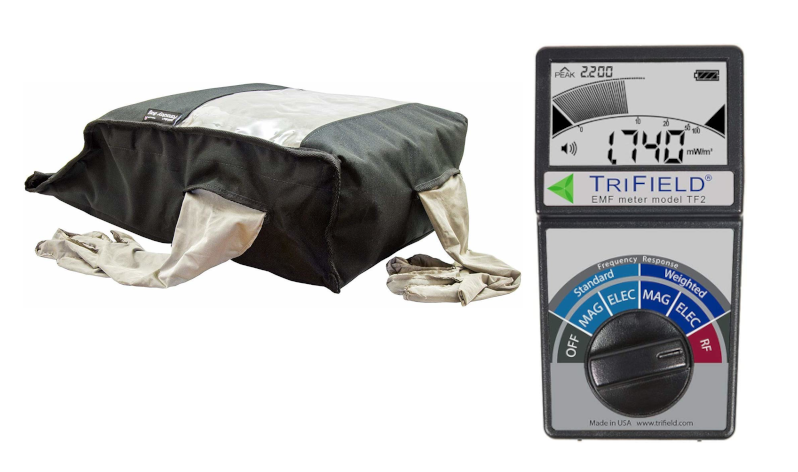}
\caption{Environment control components -- Disklabs Faraday Lab Box LB2 (leak protection), and Trifield EMF Meter Model TF2 (leak detection)}
\label{fig:lab_gear2}
\end{figure}

During the development and testing of such labs and platforms, best practices are advised: 
\begin{itemize}
\item Whenever possible or applicable, configure the transmitters (e.g., HackRF) and receivers (e.g., RTL-SDR) to use the ISM-band, meaning that the transmission and reception of the signal waves were done on the central carrier frequency of 433.800\,MHz. 
For example, in the authors' geography, the 432--438\,MHz ISM-band is allocated for transceivers exempt from licensing~\cite{traficom2021}, and it is a good practice to familiarize with the local/national regulations. 

\item Whenever possible or applicable, set the lowest transmit power to limit unintended interference in the unlicensed ISM band. 

\item In addition, use a certified ``faraday cage'' --- specifically a Disklabs Faraday Bag --- featuring a double layer military-grade RF faraday shielding, which is also commonly used for well-contained wireless and RF testing and forensics (Figure~\ref{fig:lab_gear2}). 

\item Moreover, use a certified radio power density meter --- specifically a TriField Model TF2 EMF Meter --- to double-check and ensure that the signals do not escape the faraday cage/lab premises (Figure~\ref{fig:lab_gear2}). 
\end{itemize}
All these precaution measures are complementary and ensure a well-controlled environment, which is also in line with commonly accepted practices. 


\subsection{Summary and comparison with related work}

In Table~\ref{tab:rel-work-summary}, we present a comparison of the main related work and our present paper, and below, we introduce the meaning of symbols used in Table~\ref{tab:rel-work-summary}.

\begin{itemize}

\item \includegraphics[width = 0.05\columnwidth, trim = {0.0in 0.1in 0.0in 0.0in}]{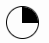}: Some of these apply to system/setup: demonstrated minor early-stage results; implementation is very early-stage; qualifies for low Technical Readiness Levels (TRL). 

\item \includegraphics[width = 0.05\columnwidth, trim = {0.0in 0.1in 0.0in 0.0in}]{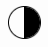}: Some of these apply to system/setup: demonstrated some limited results; implementation is partial or does not cover all use-cases; qualifies for medium Technical Readiness Levels (TRL). 

\item \includegraphics[width = 0.05\columnwidth, trim = {0.0in 0.1in 0.0in 0.0in}]{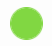}: Any (or generally all) of these apply to system/setup: generally covers all mentioned use-cases; has extensible and/or close-to-complete implementation; demonstrated extensive results; qualifies for high Technical Readiness Levels (TRL). 

\end{itemize}

\begin{table*}[htb]
\centering
\caption{Summary comparison with related State of the Art ``lab setup'' works.}
\label{tab:rel-work-summary}
\begin{tabular}{|l|c|c|c|c|c|c|}


\hline
Paper &
Satellite + Space + Aerospace &
Aviation &
Maritime &
Drones &
GPS &
RF shielding 
\\ 


\hline
Strohmeier et al.~\cite{strohmeier2022building} 

& \begin{tabular}[c]{@{}c@{}}\includegraphics[width = 0.05\columnwidth, trim = {0.0in 0.1in 0.in 0.0in}]{figures/half.png}\\ (SATCOM)\end{tabular} 

& \begin{tabular}[c]{@{}c@{}}\includegraphics[width = 0.05\columnwidth, trim = {0.0in 0.1in 0.in 0.0in}]{figures/green.png}\\ (ADS-B, TCAS, CPDLC, extensible)\end{tabular} 

&  NO

& \begin{tabular}[c]{@{}c@{}}\includegraphics[width = 0.05\columnwidth, trim = {0.0in 0.1in 0.in 0.0in}]{figures/quater.png}\\ (FLARM)\end{tabular} 

& \includegraphics[width = 0.05\columnwidth, trim = {0.0in 0.1in 0.0in 0.0in}]{figures/half.png} 

& \includegraphics[width = 0.05\columnwidth, trim = {0.0in 0.1in 0.0in 0.0in}]{figures/half.png} 
\\


\hline
Predescu et al.~\cite{Predescu2022testbed}  

&  NO

& \begin{tabular}[c]{@{}c@{}}\includegraphics[width = 0.05\columnwidth, trim = {0.0in 0.1in 0.0in 0.0in}]{figures/half.png}\\ (ARINC 429, ARINC 664)\end{tabular}

& NO 
 
& NO 
 
& NO 

& NO 
\\


\hline
Dey et al.~\cite{dey2018drone}  

& NO

& NO 

& NO
 
& \begin{tabular}[c]{@{}c@{}}\includegraphics[width = 0.05\columnwidth, trim = {0.0in 0.1in 0.0in 0.0in}]{figures/half.png} \end{tabular}
 
& \includegraphics[width = 0.05\columnwidth, trim = {0.0in 0.1in 0.0in 0.0in}]{figures/quater.png} 

& NO 
\\


\hline
\textbf{Our current paper} 

& \begin{tabular}[c]{@{}c@{}}\includegraphics[width = 0.05\columnwidth, trim = {0.0in 0.1in 0.0in 0.0in}]{figures/green.png}\\ (CCSDS, COSPAS-SARSAT, EPIRB \\ extensible + cross-channel (XC))\end{tabular}

& \begin{tabular}[c]{@{}c@{}}\includegraphics[width = 0.05\columnwidth, trim = {0.0in 0.1in 0.0in 0.0in}]{figures/green.png}\\ (ADS-B, EFB, ACARS, EPIRB-ELL \\ extensible + cross-channel (XC))\end{tabular}

& \begin{tabular}[c]{@{}c@{}}\includegraphics[width = 0.05\columnwidth, trim = {0.0in 0.1in 0.0in 0.0in}]{figures/green.png}\\ (AIS, EPIRB-MMSI \\ extensible + cross-channel (XC))\end{tabular}

& \begin{tabular}[c]{@{}c@{}}\includegraphics[width = 0.05\columnwidth, trim = {0.0in 0.1in 0.0in 0.0in}]{figures/quater.png}\\ (RemoteID, FLARM)\end{tabular}

& NO 

& \includegraphics[width = 0.05\columnwidth, trim = {0.0in 0.1in 0.0in 0.0in}]{figures/quater.png}
\\

\hline


\end{tabular}%
\end{table*}


\section{Results}
\label{sec:results}

We formulated and tested many existing and novel attacks on ADS-B and AIS with the mentioned setup. Because our setup supports encoding raw data, besides different attacks, we also sampled some technical limitations of the receivers, such as error handling capability.
%
All the experiments have been conducted within a controlled lab environment, running at minimal power and shortest duration possible.
We briefly describe the test result below.

\subsection{Experiments on satellite systems}
\label{sec:results-sat}

We conducted the following tests on the satellite and COSPAS-SARSAT EPIRB system:
\begin{itemize}
  \item Replaying
  \item Spoofing
  \item Fuzzing
  \item Denial-of-service (DoS)
\end{itemize}

\begin{figure}[!htb]
\centering
\includegraphics[width=0.90\columnwidth]{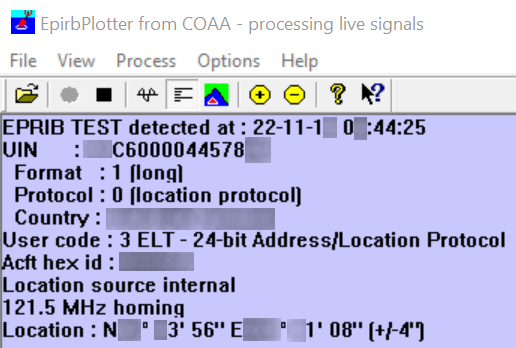}
\caption{Our \textit{spoofed} EPIRB-ELT signal (contains ICAO24 aircraft ID) well received by EpirbPlotter.}
\label{fig:elt}
\end{figure}

\begin{figure}[!htb]
\centering
\includegraphics[width=0.90\columnwidth]{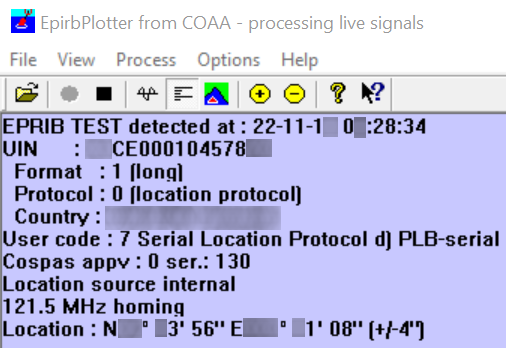}
\caption{Our \textit{spoofed} EPIRB-PLB signal well received by EpirbPlotter.}
\label{fig:plb}
\end{figure}

\begin{figure}[!htb]
\centering
\includegraphics[width=0.90\columnwidth]{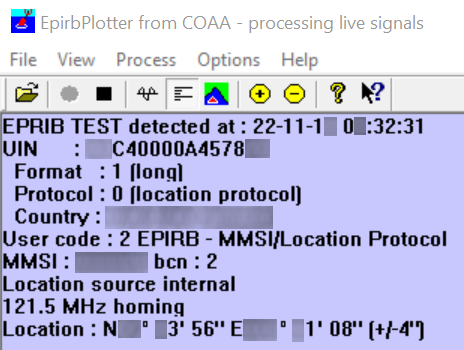}
\caption{Our \textit{spoofed} EPIRB-MMSI AIS signal (contains MMSI ship ID) well received by EpirbPlotter.}
\label{fig:epirb}
\end{figure}

On the COSPAS-SARSAT EPIRB implementations, while we were unable to achieve DoS or crashes (as only very basic EpirbPlotter software was available as the target receiver at this point), we have successfully achieved replaying, spoofing, and fuzzing. 
In Figures~\ref{fig:elt}~\ref{fig:plb}~\ref{fig:epirb}, we show successful EPIRB spoofing attacks, wehere we can accurately control virtually any field of the EPIRB messages. 

On the Theia Space ESAT, all attacks above were implemented successfully on the satellite's CCSDS implementation. Moreover, thanks to fuzzing, we discovered specific packets and CCSDS sequences that consistently trigger a quasi-permanent DoS, i.e., the device requires a hard reset in order for the communication with the device to be possible again.

One main challenge limiting the number of attacks tested is the highly-limited access to the COSPAS-SARSAT systems (including software and devices) which itself is due to either high costs or restricted access (related to the sensitive nature of such systems). 
As immediate future work, we aim to establish national and international contact points with COSPAS-SARSAT centers to bootstrap cybersecurity readiness testing and exercises involving presented and future/novel attacks.

\subsection{Experiments on avionics system}

We have successfully tested the following attacks on the ADS-B system: 
\begin{itemize}
  \item Aircraft reconnaissance
  \item Spoofing
  \item Flooding
  \item Jamming
  \item False emergency signal
  \item Aircraft disappearance
  \item Trajectory Modification
  \item Logically invalid data encoding 
  \item Fuzzing avionics protocol (GDL-90)
  \item Denial-of-Service (DoS) 
  \item Attacks on ADS-B CRC error handling
  \item Highly-Coordinated attackers attack
\end{itemize}
We implemented and tested these 12 cyberattacks on ADS-B, of which five attacks were presented or implemented for the first time. 
Six portable mobile cockpit information system (MCIS) devices combined with 21 EFBs, resulting in 44 ADS-B 1090ES and 24 UAT978 configurations, were tested for the DoS attack, which affected approximately 63\% and 37\% of 1090ES and UAT978 setups, respectively.
Moreover, the GDL 90 fuzzing experiment shows a worrying and critical lack of security in several electronic flight bag (EFB) applications. Out of 16 tested configurations, nine (56\%) were impacted (crash, hang, and abnormal behavior).

\subsection{Experiments on maritime system}
We conducted the following tests on the AIS system:
\begin{itemize}
  \item Spoofing
  \item Fake alert ``Man OverBoard'' (MOB)
  \item Fake alert ``Vessel Collision''
  \item Jamming
  \item Overwhelming alerts
  \item Visual navigation disruption
  \item Logically invalid data encoding 
  \item Denial-of-Service (DoS)
  \item Highly-Coordinated attackers attack
  \item Error handling test
  \item AIS preamble test
\end{itemize}
We implemented and tested 11 different tests/attacks on 19 AIS setups. The results showed that approximately 89\% of the setups were affected by DoS attacks. We also identified an implementation/specification flaw related to the AIS preamble during the experiments, which may affect the interoperability of different AIS devices.

\subsection{Other results and applications}

We have also experimented with expanding further application horizons of our lab. 
For example, we have been successful in using our lab's pentesting platform and its flexible capabilities to research, implement and demonstrate the effectiveness of multiple infamous \textit{log4j exploits}~\cite{hiesgen2022race} (Remote Code Execution, Denial of Service) when vulnerable components are used within aviation (ACARS, ADS-B) and maritime (AIS) infrastructure.



\section{Conclusion}
\label{sec:concl}

We presented our \emph{``Unified Cybersecurity Testing Lab for Satellite, Aerospace, Avionics, Maritime, Drone (SAAMD) technologies and communications''} -- a cybersecurity-focused, research-oriented, and industry-capable lab featuring a flexible pentesting, attack and evaluation platform. The lab aims at offering extensive and extensible capabilities to perform complex cybersecurity analyses and tests that are otherwise challenging to perform in the real world or in similar yet domain-constrained labs. 

In particular, our lab and the vision behind it bridges the space and satellite technologies with the aviation, maritime, and drone technologies and protocols, thus allowing new types and levels of research, experimentation, and innovation to be performed in a unique and highly unified manner, both for cybersecurity and non-cybersecurity purposes. 
%

Last but not least, we invite all interested researchers and industry practitioners in these domains to elaborate their novel and experimental ideas to achieve extensive collaborations and expand the utility of the lab to its maximum potential. 
All such comments, requests and queries are welcome at 
\href{mailto:ancostin@jyu.fi}{ancostin@jyu.fi}.


\section*{Acknowledgments}
\label{sec:ack}

Minor sections and some hardware of this research were kindly supported by the cascade funding from 
Engage KTN (SESAR Joint Undertaking under the European Union's Horizon 2020 research and innovation programme under grant agreement No 783287) 
project \emph{"Engage - 204 - Proof-of-concept: practical, flexible, affordable pentesting platform for ATM/avionics cybersecurity"}. 
All and any results, views, opinions are authors' only and do not reflect the official position of the European Union (and it's organizations and projects, including Horizon 2020 program and Engage KTN). 
Major parts of this research were supported by 
\emph{``Decision of the Research Dean on research funding (20.04.2022)''} 
within the Faculty of Information Technology of University of Jyv\"{a}skyl\"{a}. 
Hannu Turtiainen also thanks the Finnish Cultural Foundation / Suomen Kulttuurirahasto (https://skr.fi/en) 
for supporting his Ph.D. dissertation work and research (under grant decision no. 00221059) and the Faculty of Information Technology of the University of Jyv\"{a}skyl\"{a} (JYU), 
in particular, Prof. Timo H\"{a}m\"{a}l\"{a}inen, for partly supporting and supervising his Ph.D. work at JYU in 2021--2023.
Syed Khandker was partially supported by the Finnish Foundation for Technology Promotion under the PoDoCo grant program.




\tiny{
\bibliographystyle{IEEEtran}
\bibliography{IEEEexample.bib}
}

\end{document}